\documentclass[12pt, a4paper]{article}
\usepackage[utf8]{inputenc}
\usepackage{hyperref}
\usepackage{geometry}
\usepackage{graphicx}
\usepackage{subcaption}
\usepackage[numbers]{natbib}
\usepackage{amssymb}                                                                                                                                                                                                                                                                                                                                                                                                                                                                                                                                                                                                                                                                                                                                                                                                                                                                                                                                                                                                                                                                                                                                                                                                                                                                                                                                                                                                                                                                                                                                                                                                                                                                                                                                                                                                                                                                                                                                                                                                                                                                                                                                                                                                                                                                                                                                                                                                                                                                                                                                                                                                                                                                                                                                                                                                                                                                                                                                                                                                                                                                                                                                                                                                                                                                                                                                                                                                                                                                                                                                                                                                                                                                                                                                                                                                                                                                                                                                                                                                                                                                                                                                                                                                                                          
\usepackage{amsmath}
\usepackage{slashed}
\usepackage{anyfontsize}
\usepackage{bm}
\usepackage{slashed}
\usepackage{multirow}
\usepackage{float}
\usepackage[font=small]{caption}
\usepackage{listings}
\newgeometry{lmargin=1.1in, rmargin=1.1in, tmargin=1in, bmargin=1in}
\title{Generalized Yang-Mills Theory under Rotor Mechanism}
\author{B.T.T.Wong\footnote{CERN, u3500478@connect.hku.hk}}
\date{}

\begin{document}

\maketitle
\begin{abstract}
This paper follows the previous work on generalized abelian gauge field theory of higher-order derivatives under rotor model and extends the study to the most generalized non-abelian case. We find that the rotor mechanism from the abelian case applies nicely to the non-abelian case under the Lorentz gauge condition.  Under the rotor mechanism, the gauge field transforms as $T_{\mu}^a \rightarrow \Box^n T_{\mu}^a$. When the order of field derivative is $n=0$, this restores back to the original Yang-Mills action. Our work gives an extensive generalization of the Yang-Mills theory with higher-order field derivatives. We also compute the equation of motion and Noether’s current of the generalized non-abelian gauge field theory. Finally, we study the dynamic instability issue of the theory by the Ostrogradsky construction and the analysis of the 00-component of the energy-momentum tensor.
 
\end{abstract}

\section{Introduction}
The study of high-order derivatives system has a long history dated back to 1950s, with the classic development of Pais-Uhlenbeck oscillator \cite{first} and Podolsky electrodynamics \cite{ho1,ho2,ho3,ho4}. Quantum field theories with high-order derivatives are of appealing interest because they have the possibilities to eliminate ultraviolet divergences in the calculation of scattering amplitudes \cite{ho1, ho2, ho3, ho4, ho5, LeeWick1, LeeWick2, LeeWickSM}. Vast amount of studies in higher order derivative field theories including both scalar fields and gauge fields have been performed \cite{first, ho6a,ho6b,ho6c}. However, there are difficulties in establishing the formalism as these theories are often non-renormalizable and dynamically unstable with unbounded Hamiltonian \cite{ho6d, ho6e, ho6f, ho6g, ho6h}. Yet, higher derivative field theories give insight to the study of quantum gravity and modified theories of gravity and hence it is worth to establish new formalisms on high-order theories \cite{h1,h2,h3, h4,h5}.
  
In our previous paper \cite{BW}, we have established the formalism of generalized abelian gauge field theory under rotor model with higher-order derivatives. We have shown the following theorem:
\begin{equation}
S = -\frac{1}{4} \int d^D x G_{n\,\mu\nu} G^{\mu\nu}_n  = \frac{1}{4^n}\int d^D x \big( \Box^n T^{\mu} \big) \hat{R}_{\mu\nu}  \big( \Box^n T^{\nu} \big) =- \frac{1}{4^{n+1}} \int d^{D}x \,\Box^n G_{\mu \nu} \Box^n G^{\mu \nu}\,,
\end{equation}
where $\hat{R}_{\mu\nu}$ is the projection tensor $\hat{R}_{\mu\nu} = \frac{1}{2}( \Box \eta_{\mu\nu} - \partial_{\mu}\partial_{\nu}) $ and $G_{n\,\mu\nu}= \partial_{\mu}T_{n\,\nu}-\partial_{\nu}T_{n\,\mu} $ is the field strength of the $n$-th order rotor model. We have the $n$-th order gauge field strength  identified as \cite{BW}
\begin{equation}
G_{n\,\mu\nu} \equiv \frac{1}{2^n} \Box^n G_{\mu\nu} \,.
\end{equation}
The projection is regarded as the second-order rotation \cite{BW}. Therefore, the $n$-th order rotation of gauge field $T_{n\,\mu_n}$ is attained by successive second-order rotations given by \cite{BW}
\begin{equation} \label{eq:rotor}
T_{n\,\mu_n} = \hat{R}_{\mu_n \mu_{n-1}}\hat{R}^{\mu_{n-1} \mu_{n-2}} \cdots \hat{R}_{\mu_3 \mu_2} \hat{R}^{\mu_2 \mu_1} \hat{R}_{\mu_1 \mu_0} T^{\mu_0} = \frac{1}{2^{n-1}}  P_{\mu_n}^{\,\,\,\mu_{n-1}} P_{\mu_{n-1}}^{\,\,\,\mu_{n-2}} \cdots P_{\mu_3}^{\,\,\,\mu_{2}} P_{\mu_2}^{\,\,\,\mu_{1}} \hat{R}_{\mu_1 \mu_0} T^{\mu_0} \,,
\end{equation}
for which 
\begin{equation} \label{eq:propagator}
P_{\mu_j}^{\,\,\,\mu_{j-1}} = \Box \delta_{\mu_j}^{\,\,\,\mu_{j-1}}
\end{equation}
is called the propagator. This is known as the rotor transformation which generates high-order derivative gauge fields. In the generalized theory, the action changes by the transformation of gauge field as  $T^{\mu} \rightarrow \Box^n T^{\mu}$. The working dimension $D$ for renormalizability is $4n+4$ for unity gauge field dimension \cite{BW}. When $n=0$ this reduces back to the conventional Maxwell action
\begin{equation} \label{eq:AbelianGaugeAction}
S = -\frac{1}{4} \int d^4 x G_{\mu\nu} G^{\mu\nu} = \int d^4 x T^{\mu}\hat{R}_{\mu\nu} T^{\nu} \,.
\end{equation}

In this article, we aim at constructing a generalized non-abelian gauge field theory (Yang-Mills theory) under the rotor mechanism. There is no guarantee that the old mechanism applies due to the existence of extra terms in the action for the non-abelian case. For the non-abelian case, we have the Yang-Mills action in general $D$ dimensional spacetime as \citep{YangMills},
\begin{equation} \label{eq:YM}
S_{\mathrm{YM}} = -\frac{1}{2} \int d^D x \mathrm{Tr}\,G_{\mu\nu} G^{\mu\nu} = -\frac{1}{4} \int d^D x \,G^a_{\mu\nu} G^{\mu\nu\,a} \,
\end{equation}
where 
\begin{equation}
G_{\mu\nu} = \partial_{\mu}T_{\nu} -\partial_{\nu}T_{\mu} -ig[T_{\mu}, T_{\nu} ] \,,
\end{equation}
for which $G_{\mu\nu}=F^a_{\mu\nu}t^a$ and $T_{\mu} = T_{\mu}^a t^a $ are matrices with $t^a$ the generators of $\mathrm{SU}(N)$ Lie group. We have to sum over repeated generator indices $a$. Using the Lie algebra $[t^a , t^b] =if^{abc}t^c$, this gives the gauge field strength as,
\begin{equation}
G_{\mu\nu}^a = \partial_{\mu}T_{\nu}^a -\partial_{\nu}T_{\mu}^a +gf^{abc}T_{\mu}^b T_{\nu}^c\,.
\end{equation}
Using integration by parts on the kinetic term, equation(\ref{eq:YM}) gives
\begin{equation} \label{eq:basic}
S_{\mathrm{YM}} = \int d^D x  \bigg( T^{\mu a} \hat{R}_{\mu\nu} T^{\nu a} - \frac{g}{2} f^{abc}(\partial^{\mu}T^{\nu a} - \partial^{\nu}T^{\mu a})T_{\mu}^b T_{\nu}^c - \frac{g^2}{4} f^{abc}f^{ade} T_{\mu}^b T_{\nu}^c T^{\mu d} T^{\nu e} \bigg)\,,
\end{equation}
where the first term is Maxwell-like, and we have the second term and the third as the extra self-coupling terms compared to the abelian case. Let's define the Maxwell-like action as
\begin{equation}
S_{\mathrm{Maxwell}} = \int d^D x  ( T^{\mu a} \hat{R}_{\mu\nu} T^{\nu a} ) \,,
\end{equation}
in such a way that we can use all the results we have in our previous work\citep{BW}. We also define the Maxwell-like gauge field strength as $\tilde{G}_{\mu\nu} = \partial_{\mu}T_{\nu} -\partial_{\nu}T_{\mu}$, so that
\begin{equation}
G_{\mu\nu} = \tilde{G}_{\mu\nu} -ig[T_{\mu}, T_{\nu} ] \,,
\end{equation}
and
\begin{equation}
G_{\mu\nu}^a \rightarrow \tilde{G}_{\mu\nu}^a + gf^{abc}T_{\mu}^b T_{\nu}^c\,.
\end{equation}
Under the rotor mechanism, the non-abelian gauge field transforms as
\begin{equation}
T^{\mu_0 a} \rightarrow T^a_{n\,\mu_n} = (  \hat{R}_{\mu_n \mu_{n-1}}\hat{R}^{\mu_{n-1} \mu_{n-2}} \cdots \hat{R}_{\mu_3 \mu_2} \hat{R}^{\mu_2 \mu_1} \hat{R}_{\mu_1 \mu_0}  )T^{\mu_0 a} \,.
\end{equation}

\section{The generalized non-abelian gauge field theorem under rotor model}
In this paper, in analogy to the abelian case of our previous work in \citep{BW}, we aim at proving the following theorem for the non-abelian case,
\begin{equation} \label{eq:general}
\begin{aligned}
S^{(n)}_{\mathrm{YM}}&=-\frac{1}{4} \int d^D x \,G^a_{n\,\mu\nu} G^{\mu\nu\,a}_n \\
&=\int d^D x  \bigg( \frac{1}{4^n} \Box^n T^{\mu a} \hat{R}_{\mu\nu} \Box^n T^{\nu a} - \frac{1}{2\cdot 8^n}g f^{abc}(\partial^{\mu}\Box^n T^{\nu a} - \partial^{\nu} \Box^n T^{\mu a})\Box^n T_{\mu}^b \Box^n T_{\nu}^c \\
&\quad\quad \quad\quad\quad - \frac{g^2}{4\cdot 16^n} f^{abc}f^{ade} \Box^n T_{\mu}^b \Box^n T_{\nu}^c \Box^n T^{\mu d} \Box^n T^{\nu e} \bigg) 
\end{aligned}
\end{equation}
We will see that indeed the rotor mechanism in the abelian case applies to the non-abelian case as well without modification. The only necessary condition is the Lorentz gauge condition $\partial_{\mu}T^{\mu a}= 0$. In such way the rotor mechanism functions the gauge field to transform as 
\begin{equation}
T_\mu^a \rightarrow \Box^n T_\mu^a \,.
\end{equation}
And the gauge field strength transforms as
\begin{equation}
G_{\mu\nu} \rightarrow  \partial_{\mu}\Box^n T_{\nu} -\partial_{\nu}\Box^n T_{\mu} -ig[\Box^n T_{\mu}, \Box^n T_{\nu} ] = \Box^n \tilde{G}_{\mu\nu} -ig ( \Box^n T_{\mu} \Box^n T_{\nu} - \Box^n T_{\nu} \Box^n T_{\mu}) \,.
\end{equation}
or explicitly
\begin{equation}
G_{\mu\nu}^a  \rightarrow \partial_{\mu} \Box^n T_{\nu}^a -\partial_{\nu} \Box^n T_{\mu}^a + gf^{abc} \Box^n T_{\mu}^b \Box^n T_{\nu}^c = \Box^n \tilde{G}^a_{\mu\nu} + gf^{abc} \Box^n T_{\mu}^b \Box^n T_{\nu}^c\,.
\end{equation}
When $n=0$, then general form of equation (\ref{eq:general}) will reduce back to (\ref{eq:basic}). We will first prove the theorem for $n=1$, followed by $n=2$ case and finally the general $n$ case as we do in \citep{BW}.
\subsection{The $n=1$ case}
For the first-ordered rotation, the first-ordered rotated field is \begin{equation}
L_{\mu}^a = \hat{R}_{\mu\nu}T^{\nu a} \,.
\end{equation}
The new gauge field strength is
\begin{equation}
H^a_{\mu\nu} = \partial_{\mu}L_{\nu}^a -\partial_{\nu}L_{\mu}^a + gf^{abc}L_{\mu}^b L_{\nu}^c\,.
\end{equation}
The new action becomes
\begin{equation}
\begin{aligned}
S^{(1)}_{\mathrm{YM}} = \int d^D x  \bigg( L^{\mu a} \hat{R}_{\mu\nu} L^{\nu a} - \frac{g}{2} f^{abc}(\partial^{\mu}L^{\nu a} - \partial^{\nu}L^{\mu a})L_{\mu}^b L_{\nu}^c -\frac{g^2}{4} f^{abc}f^{ade} L_{\mu}^b L_{\nu}^c L^{\mu d} L^{\nu e} \bigg)\,,
\end{aligned}
\end{equation}
For the first Maxwell-like term, we can apply the result we obtained in our previous work \citep{BW} (equation 27), in which we have
\begin{equation}
\int d^D x  L^{\mu a} \hat{R}_{\mu\nu} L^{\nu a} = \frac{1}{4} \int d^D x \Box T^{\mu a} \hat{R}_{\mu\nu} \Box T^{\nu a} \,. 
\end{equation}
Then we have
\begin{equation}
\begin{aligned}
S^{(1)}_{\mathrm{YM}} &= \int d^D x  \bigg( \frac{1}{4} \Box T^{\mu a} \hat{R}_{\mu\nu} \Box T^{\nu a} - \frac{g}{2} f^{abc}(\partial^{\mu}L^{\nu a} - \partial^{\nu}L^{\mu a})L_{\mu}^b L_{\nu}^c -\frac{ g^2}{4} f^{abc}f^{ade} L_{\mu}^b L_{\nu}^c L^{\mu d} L^{\nu e} \bigg) \\
&=\int d^D x  \bigg( \frac{1}{4} \Box T^{\mu a} \hat{R}_{\mu\nu} \Box T^{\nu a}  -\frac{g}{2} f^{abc}(\partial^{\mu} \hat{R}^{\nu\delta}T^a_{\delta} - \partial^{\nu}\hat{R}^{\mu\lambda} T^a_\lambda ) \hat{R}_{\mu\alpha} T^{\alpha b} \hat{R}_{\nu\beta}T^{\beta c} \\
&\quad\quad\quad\quad\quad\quad\quad\quad\quad\quad\quad - \frac{g^2}{4} f^{abc}f^{ade}\hat{R}_{\mu\alpha}T^{\alpha b} R_{\nu\beta}T^{\beta c} R^{\mu\gamma}T^d_{\gamma}R^{\nu\delta}T^e_{\delta}\bigg)\,.
\end{aligned}
\end{equation}
Now we investigate the second term and the third term of the last line. For the second term, we have
\begin{equation}
\begin{aligned}
&-\quad \frac{g}{2} f^{abc}(\partial^{\mu} \hat{R}^{\nu\delta}T^a_{\delta} - \partial^{\nu}\hat{R}^{\mu\lambda} T^a_\lambda ) \hat{R}_{\mu\alpha} T^{\alpha b} \hat{R}_{\nu\beta}T^{\beta c} \\
&=-\frac{1}{2\cdot 8}g f^{abc}\Big( \partial^{\mu}(\Box\eta^{\nu\delta}-\partial^{\nu}\partial^{\delta})T^a_{\delta} -\partial^{\nu}(\Box\eta^{\mu\lambda}-\partial^{\mu}\partial^{\lambda})T^a_{\lambda}    \Big)(\Box\eta_{\mu\alpha}-\partial_{\mu} \partial_{\alpha})T^{\alpha b}\\
&\quad\quad\quad\quad\quad\quad\quad\quad\quad\quad\quad\quad\quad\quad\quad\quad\quad\quad\quad\quad\quad\quad\quad\quad\times(\Box\eta_{\nu\beta}-\partial_{\nu}\partial_{\beta})T^{\beta c} \\
&=-\frac{1}{16}g f^{abc}\Big( (\partial^{\mu}\Box\eta^{\nu\delta}T^a_{\delta}-\partial^{\nu}\Box\eta^{\mu\lambda}T^{a}_{\lambda} -\partial^{\mu}\partial^{\nu}\partial^{\delta}T^a_{\delta} + \partial^{\nu}\partial^{\mu}\partial^{\lambda}T^a_{\lambda}) \\
&\quad\quad\quad\quad\quad\quad\quad\quad \times( \Box T_{\mu}^b \Box T^c_{\nu} - \Box T^b_{\mu}\partial_{\nu} \partial_{\beta} T^{\beta c} - \partial_{\mu}\partial_{\alpha}T^{\alpha b}\Box T_{\nu}^c +(\partial_{\mu}\partial_{\alpha} T^{\alpha b})(\partial_{\nu}\partial_{\beta}T^{\beta c})) \Big) \\
&= -\frac{1}{16}g f^{abc}\Big( (\partial^{\mu}\Box\eta^{\nu\delta}T^a_{\delta}-\partial^{\nu}\Box\eta^{\mu\lambda}T^{a}_{\lambda} ) \\
&\quad\quad\quad\quad\quad\quad\quad\quad \times(\Box T_{\mu}^b \Box T^c_{\nu} - \Box T^b_{\mu}\partial_{\nu}\partial_{\beta} T^{\beta c} - \partial_{\mu}\partial_{\alpha}T^{\alpha b}\Box T_{\nu}^c +(\partial_{\mu}\partial_{\alpha} T^{\alpha b})(\partial_{\nu}\partial_{\beta}T^{\beta c})) \Big) \,. \\
\end{aligned}
\end{equation}
Now we impose the Lorentz gauge condition, then the last three terms vanish. Therefore we have
\begin{equation}
\begin{aligned}
&\quad -\frac{g}{2} f^{abc}(\partial^{\mu} \hat{R}^{\nu\delta}T^a_{\delta} - \partial^{\nu}\hat{R}^{\mu\lambda} T^a_\lambda ) \hat{R}_{\mu\alpha} T^{\alpha b} \hat{R}_{\nu\beta}T^{\beta c} \\
&=-\frac{1}{16}g f^{abc}(\partial^{\mu}\Box T^{\nu a}-\partial^{\nu}\Box T^{\mu a} )\Box T_{\mu}^b \Box T^c_{\nu} \,.
\end{aligned}
\end{equation}
For the third term, we would also impose the Lorentz gauge condition,
\begin{equation}
\begin{aligned}
&\quad -\frac{g^2}{4} f^{abc}f^{ade}\hat{R}_{\mu\alpha}T^{\alpha b} R_{\nu\beta}T^{\beta c} R^{\mu\gamma}T^d_{\gamma}R^{\nu\delta}T^e_{\delta} \\
&=  -\frac{g^2}{64}f^{abc}f^{ade}(\Box\eta_{\mu\alpha}T^{\alpha b}-\partial_{\mu}\partial_{\alpha}T^{\alpha b} )( \Box\eta_{\nu\beta}T^{\beta c}-\partial_{\nu}\partial_{\beta}T^{\beta c}) \\
&\quad\quad\quad\quad\quad\quad\quad\quad\quad\quad\quad\quad \times ( \Box\eta^{\mu\gamma}T^{d}_{\gamma}-\partial^{\mu}\partial^{\gamma}T^{d}_\gamma )( \Box\eta^{\nu\delta}T^{e}_{\delta}-\partial^{\nu}\partial^{\delta}T^{e}_\delta )\\
& =-\frac{g^2}{64}f^{abc}f^{ade} \Box T^b_{\mu} \Box T^c_{\nu} \Box T^{\mu d} \Box T^{\nu e} \,.
\end{aligned}
\end{equation}
Therefore, for the $n=1$ case, we obtain
\begin{equation}
S^{(1)}_{\mathrm{YM}} = \int d^D x  \bigg( \frac{1}{4} \Box T^{\mu a} \hat{R}_{\mu\nu} \Box T^{\nu a} -  \frac{g}{16} f^{abc}(\partial^{\mu}\Box T^{\nu a}-\partial^{\nu}\Box T^{\mu a} )\Box T_{\mu}^b \Box T^c_{\nu} -\frac{g^2}{64}f^{abc}f^{ade} \Box T^b_{\mu} \Box T^c_{\nu} \Box T^{\mu d} \Box T^{\nu e} \bigg)\,.
\end{equation}
Thus the proof is completed.

\subsection{The $n=2$ case}
For the second-ordered rotation, the second-ordered rotation field is
\begin{equation}
J^{\rho a}=R^{\rho\mu}R_{\mu\nu}T^{\nu a}\,.
\end{equation}
Using the result in \citep{BW} (equation 30), 
\begin{equation}
J^{\mu a} = \frac{1}{4}(\Box^2 \delta_{\alpha}^{\mu} - \Box \partial^{\mu}\partial_{\alpha}  ) T^{\alpha a}\,.
\end{equation}
Also we have
\begin{equation}
J^a_{\sigma}=\eta_{\sigma\rho}\hat{R}^{\rho\mu}\hat{R}_{\mu\nu}T^{\nu a} =\frac{1}{4}(\Box^2 \eta_{\sigma\alpha} -\Box\partial_{\sigma}\partial_{\alpha} )T^{\alpha a} \,.
\end{equation}
The new gauge field strength is
\begin{equation}
K^a_{\mu\nu} = \partial_{\mu}J_{\nu}^a -\partial_{\nu}J_{\mu}^a + gf^{abc}J_{\mu}^b J_{\nu}^c\,.
\end{equation}
The new action becomes
\begin{equation}
\begin{aligned}
S^{(2)}_{\mathrm{YM}} = \int d^D x  \bigg( J^{\mu a} \hat{R}_{\mu\nu} J^{\nu a} - \frac{g}{2} f^{abc}(\partial^{\mu}J^{\nu a} - \partial^{\nu}L^{\mu a})J_{\mu}^b J_{\nu}^c - \frac{g^2}{4} f^{abc}f^{ade} J_{\mu}^b J_{\nu}^c J^{\mu d} J^{\nu e} \bigg)\,,
\end{aligned}
\end{equation}
For the first Maxwell-like term, we can apply the result we obtained in our previous paper \citep{BW} (equation (43)), in which we have,
\begin{equation}
\int d^D x  J^{\mu a} \hat{R}_{\mu\nu} J^{\nu a} = \frac{1}{16} \int d^D x \Box^2 T^{\mu a} \hat{R}_{\mu\nu} \Box^2 T^{\nu a} \,. 
\end{equation}
Explicitly, this is
\begin{equation}
\begin{aligned}
S^{(2)}_{\mathrm{YM}} &= \int d^D x  \bigg(\frac{1}{16} \Box^2 T^{\mu a} \hat{R}_{\mu\nu} \Box^2 T^{\nu a}- \frac{g}{2} f^{abc}(\partial^\mu\hat{R}^{\nu\rho}\hat{R}_{\rho\alpha}T^{\alpha a}-\partial^\nu \hat{R}^{\mu\rho}\hat{R}_{\rho\alpha}T^{\alpha a}   ) \\
&\quad\quad\quad\quad\quad\quad\quad\quad\quad\quad\quad\quad\quad\quad\quad\quad\quad\times \eta_{\mu\rho}\eta_{\nu\kappa}\hat{R}^{\rho\sigma}\hat{R}_{\sigma\alpha}T^{\alpha b} \hat{R}^{\kappa\epsilon}\hat{R}_{\epsilon\beta}T^{\beta b} \\
&\quad\quad\quad\quad\quad\quad\quad - \frac{g^2}{4} f^{abc}f^{ade}\eta_{\mu\rho}\eta_{\nu\kappa}\hat{R}^{\rho\sigma}\hat{R}_{\sigma\alpha}T^{\alpha b} \hat{R}^{\kappa\epsilon}\hat{R}_{\epsilon\beta}T^{\beta b}\hat{R}^{\mu\delta}\hat{R}_{\delta\gamma}T^{\gamma d} \hat{R}^{\nu\tau}\hat{R}_{\tau\lambda}T^{\lambda e} \bigg)\,.
\end{aligned}
\end{equation}
Then we have
\begin{equation}
\begin{aligned}
S^{(2)}_{\mathrm{YM}} &=\int d^D x \bigg( \frac{1}{16}  \Box^2 T^{\mu a} \hat{R}_{\mu\nu} \Box^2 T^{\nu a} - \frac{1}{128}gf^{abc}\bigg(\Big( \partial^\mu (\Box^2 \delta^{\nu}_{\alpha} -\Box\partial^{\nu}\partial_{\alpha} )T^{\alpha a}- \partial^\nu (\Box^2 \delta^{\mu}_{\alpha} -\Box\partial^{\mu}\partial_{\alpha} )T^{\alpha a} \Big)\\
& \quad \times( \Box^2 \eta_{\mu\alpha} T^{\alpha b} -\Box\partial_{\mu}\partial_{\alpha} T^{\alpha b}  ) ( \Box^2 \eta_{\nu\beta} T^{\beta b} -\Box\partial_{\nu}\partial_{\beta} T^{\beta b}   ) \\
&\quad -\frac{g^2}{1024}(\Box^2 \eta_{\mu\alpha}T^{\alpha b}-\Box\partial_{\mu}\partial_{\alpha}T^{\alpha b} )(\Box^2 \eta_{\nu\beta}T^{\beta c} -\Box\partial_{\nu}\partial_{\beta}T^{\beta c}   )(\Box^2 \delta^\mu_{\gamma}T^{\gamma d}-\Box \partial^{\mu}\partial_{\gamma}T^{\gamma d}   )\\
& \quad \quad \quad \quad \quad \quad \quad \quad \quad \quad \quad \quad \quad \quad \quad \quad \quad\quad \quad \quad\quad \quad \quad\quad \quad \quad \times(\Box^2 \delta^{\nu}_{\lambda}T^{\lambda e}-\Box\partial^{\nu}\partial_{\lambda}T^{\lambda e}   )\bigg) \,.
\end{aligned}
\end{equation}
Applying Lorentz gauge condition, we obtain
\begin{equation}
\begin{aligned}
S^{(2)}_{\mathrm{YM}} &=\int d^D x \bigg( \frac{1}{16}  \Box^2 T^{\mu a} \hat{R}_{\mu\nu} \Box^2 T^{\nu a} - \frac{1}{128}gf^{abc}\Big( \partial^\mu \Box^2 \delta^{\nu}_{\alpha} - \partial^\nu \Box^2 \delta^{\mu}_{\alpha} \Big)T^{\alpha a} \Box^2 \eta_{\mu\alpha} T^{\alpha b} \Box^2 \eta_{\nu\beta} T^{\beta b} \\
&\quad\quad\quad\quad\quad -\frac{g^2}{1024}\Box^2 \eta_{\mu\alpha}T^{\alpha b}\Box^2 \eta_{\nu\beta}T^{\beta c} \Box^2 \delta^\mu_{\gamma}T^{\gamma d}
\Box^2 \delta^{\nu}_{\lambda}T^{\lambda e} \bigg)\\
&=\int d^D x \bigg( \frac{1}{16}  \Box^2 T^{\mu a} \hat{R}_{\mu\nu} \Box^2 T^{\nu a}- \frac{1}{128}gf^{abc}\Big( \partial^\mu \Box^2 T^{\nu a} - \partial^\nu \Box^2 T^{\mu a} \Big) \Box^2  T^{ b}_{\mu} \Box^2  T^{b}_{\nu} \\
&\quad\quad\quad\quad\quad -\frac{g^2}{1024}\Box^2 T^{b}_\mu \Box^2 T^{c}_\nu \Box^2 T^{\mu d} \Box^2 T^{\nu e} \bigg) \,.
\end{aligned}
\end{equation}
which completes the proof for $n=2$ case.

\section{The general $n$ case}
To prove the general $n$ case, we use the results in our previous paper \citep{BW} for $n-$th order rotation and apply to the non-abelian case. Equation (\ref{eq:rotor}) and equation give (\ref{eq:propagator})
\begin{equation}
T^a_{n\,\mu_n} = (  \hat{R}_{\mu_n \mu_{n-1}}\hat{R}^{\mu_{n-1} \mu_{n-2}} \cdots \hat{R}_{\mu_3 \mu_2} \hat{R}^{\mu_2 \mu_1} \hat{R}_{\mu_1 \mu_0}  )T^{\mu_0 a} = \frac{1}{2^{n-1}}\Big(\Box^{n-1} \delta_{\mu_n}^{\,\,\,\mu_1} \Big)\hat{R}_{\mu_1 \mu_0}T^{\mu_0 a} \,, 
\end{equation}
The new gauge field strength is
\begin{equation}
G^a_{n\,\mu\nu} = \partial_{\mu}T_{n\,\nu}^a -\partial_{\nu}T_{n\,\mu}^a + gf^{abc}T_{n\,\mu}^b T_{n\,\nu}^c\,.
\end{equation}
The new action becomes
\begin{equation}
\begin{aligned}
S^{(2)}_{\mathrm{YM}} = \int d^D x  \bigg( T^{\mu a}_n \hat{R}_{\mu\nu} T^{\nu a}_n - \frac{g}{2} f^{abc}(\partial^{\mu}T^{\nu a}_n - \partial^{\nu}T^{\mu a}_n)T_{n\,\mu}^b T_{n\,\nu}^c - \frac{g^2}{4} f^{abc}f^{ade} T_{n\,\mu}^b T_{n\,\nu}^c T^{\mu d}_n T^{\nu e}_n \bigg)\,,
\end{aligned}
\end{equation}
For the first Maxwell-like term, we can apply the result we obtained in our previous paper \citep{BW} (equation (75)), in which we have,
\begin{equation}
\int d^D x T^{\mu a}_n \hat{R}_{\mu\nu} T^{\nu a}_n = \frac{1}{4^n} \int d^D x \Box^n T^{\mu a} \hat{R}_{\mu\nu} \Box^n T^{\nu a} \,. 
\end{equation}
Therefore the action is
\begin{equation}
\begin{aligned}
S^{(2)}_{\mathrm{YM}} &=\int d^D x  \bigg(\frac{1}{4^n} \Box^n T^{\mu a} \hat{R}_{\mu\nu} \Box^n T^{\nu a} - \frac{1}{2^{3n+1}}g f^{abc}(\partial^\mu \Box^{n-1}    \eta^{\nu\mu_1}\hat{R}_{\mu_1 \mu_0}T^{\mu_0 a} \\
&\quad\quad - \partial^\nu \Box^{n-1}\eta^{\mu\mu_1}\hat{R}_{\mu_1 \mu_0}T^{\mu_0 a})(\Box^{n-1} \delta_{\mu}^{\,\,\rho_1}\hat{R}_{\rho_1 \beta}T^{\beta b} )  (\Box^{n-1}\delta_\nu^{\,\,\sigma_1}\hat{R}_{\sigma_1 \gamma} T^{\gamma c} ) \\
&\quad\quad- \frac{g^2}{2^{4n+2}}f^{abc}f^{ade}(\Box^{n-1}\delta_{\mu}^{\,\,\mu_1} \hat{R}_{\mu_1 \mu_0} T^{\mu_0 b} ) (\Box^{n-1}\delta_{\nu}^{\,\,\alpha_1} \hat{R}_{\alpha_1 \alpha_0} T^{\alpha_0 c} ) (\Box^{n-1}\eta^{\mu\beta_1} \hat{R}_{\beta_1 \beta_0} T^{\beta_0 d} )\\
& \quad\quad\quad\quad\quad\quad\quad\quad\quad\quad\quad\quad\quad\quad\quad\quad\quad\quad\quad\quad\quad\quad\times (\Box^{n-1}\eta^{\nu\gamma_1} \hat{R}_{\gamma_1 \gamma_0} T^{\gamma_0 e} )\bigg) \,.
\end{aligned}
\end{equation}
But since under the Lorentz gauge, we immediately have
\begin{equation}
T_{n\,\mu_n} =\frac{1}{2^{n-1}}\Big(\Box^{n-1} \delta_{\mu_n}^{\,\,\,\mu_1} \Big)\cdot\frac{1}{2}(\Box \eta_{\mu_1 \mu_0} -\partial_{\mu_1}\partial_{\mu_0})T^{\mu_0}= \frac{1}{2^n}\Box^n T_{\mu_n}\,. 
\end{equation}
Therefore we have
\begin{equation} \label{eq:transform}
\begin{aligned}
S^{(n)}_{\mathrm{YM}}&=-\frac{1}{4} \int d^D x \,G^a_{n\,\mu\nu} G^{\mu\nu\,a}_n \\
&=\int d^D x  \bigg( \frac{1}{4^n} \Box^n T^{\mu a} \hat{R}_{\mu\nu} \Box^n T^{\nu a} - \frac{1}{2\cdot 8^n }g f^{abc}(\partial^{\mu}\Box^n T^{\nu a} - \partial^{\nu} \Box^n T^{\mu a})\Box^n T_{\mu}^b \Box^n T_{\nu}^c \\
&\quad\quad \quad\quad\quad - \frac{g^2}{4\cdot 16^n} f^{abc}f^{ade} \Box^n T_{\mu}^b \Box^n T_{\nu}^c \Box^n T^{\mu d} \Box^n T^{\nu e} \bigg) \,,
\end{aligned}
\end{equation}
which completes the proof of the general $n$ case. Therefore, under the rotor mechanism, the non-abelian gauge field transforms as $T_{\mu}^a \rightarrow \Box^n T_{\mu}^a$ as we can compare to the original action in equation (\ref{eq:basic}). When $n=0$, this would give us back the original Yang-Mills action. The remormalization dimension that requires to keep unity dimension of $T_{\mu}^a$ field would be same as the abelian case, which is is $D=4n+4$ \citep{BW}. 

\section{Equation of motion and Noether’s current of the generalized
abelian gauge field theory}
The equation of motion of the original Yang-Mill's theory is given by \citep{YangMills, Peskin},
\begin{equation} \label{eq:EOM}
\partial_{\mu} F^{\mu\nu a} + gf^{abc}T_{\mu}^b G^{\mu\nu c} = 0 \,.
\end{equation}
The covariant derivative is defined by
\begin{equation}
D_{\mu} = \partial_{\mu} - ig T_{\mu}^c t^c \,.
\end{equation}
In tensor form,
\begin{equation}
D_{\mu}^{ab} = \partial_{\mu}\delta^{ab} - ig T_{\mu}^c (t^c)^{ab} \,.
\end{equation}
Using the adjoint representation $(t^k)^{ij} = -if^{kij}$, we would have
\begin{equation}
D_{\mu}^{ab}  =\partial_{\mu}\delta^{ab} + gf^{acb}T_{\mu}^c  \,.
\end{equation}
Therefore, the equation of motion in (\ref{eq:EOM}) can be neatly written by the covariant derivative acting on the non-abelian gauge field strength,
\begin{equation} \label{eq:EOMb}
D_{\mu} G^{\mu \nu a} = 0 \,.
\end{equation}
Expanding (\ref{eq:EOM}) or (\ref{eq:EOMb}) explicitly gives 
\begin{equation} \label{eq:expandEOM}
2\hat{R}^{\mu\nu}T_\mu^a +gf^{abc} (2T_{\mu}^a \partial^{\mu}T^{\nu c} + (\partial_{\mu} T^{\mu b})T^{\nu c} +T_{\mu}^b \partial^\nu T^{\mu k} + gf^{cde}T_{\mu}^b T^{\mu d}T^{\nu e}) =0\,.
\end{equation}
Under Lorentz gauge, the equation of motion simplifies to
\begin{equation}
2\Box T^{\nu a} +gf^{abc} (2T_{\mu}^a \partial^{\mu}T^{\nu c} +T_{\mu}^b \partial^\nu T^{\mu k} + gf^{cde}T_{\mu}^b T^{\mu d}T^{\nu e}) =0\,.
\end{equation}
When we compare it to the equation of motion of the abelian case $2\hat{R}^{\mu\nu}T_\mu =0$, we see that the non-abelian case contains extra non-linear terms.

To find the equation of motion for our generalized rotor model for non-abelian case, we need to vary the action as follow,
\begin{equation} \label{eq:action}
\delta S_{\mathrm{YM}}^{(n)} = -\frac{1}{2} \int d^D x G_n^{\mu\nu a} \delta G_{n\,\mu\nu}^a \,.
\end{equation} 
The remaining work is to compute $\delta G_{n\,\mu\nu}^a$. Using the transformation of field under the Lorentz gauge as in (\ref{eq:transform}),
\begin{equation}
T_{ n \,\mu} = \frac{1}{2^n} \Box^n T_\mu \,,
\end{equation}
first we have,
\begin{equation}
G_{n\mu\nu}^a = \frac{1}{2^n} \partial_{\mu} \Box^n T_{\nu}^{ a} - \frac{1}{2^n} \partial_{\nu} \Box^n T_{\mu}^{ a} + \frac{1}{4^n} gf^{abc} \Box^n T^b_\mu  \Box^n T^c_\nu\,.
\end{equation}
Then we have the variation as
\begin{equation}
\begin{aligned}
\delta G_{n\,\mu\nu}^a &= \frac{1}{2^n} \partial_{\mu} \delta\Box^n T_{\nu}^{ a} - \frac{1}{2^n} \partial_{\nu} \delta \Box^n T_{\mu}^{ a} + \frac{1}{4^n} gf^{abc} ( \Box^n T^b_\mu \delta \Box^n T^c_\nu  +  \delta\Box^n T^b_\mu  \Box^n T^c_\nu )\\
&= \frac{1}{2^n} \partial_{\mu} \delta\Box^n T_{\nu}^{ a} - \frac{1}{2^n} \partial_{\nu} \delta \Box^n T_{\mu}^{ a} + \frac{1}{4^n} gf^{abc}  \Box^n T^b_\mu  \delta\Box^n T^c_\nu  +  \frac{1}{4^n} gf^{abc} \Box^n T^c_\nu \delta\Box^n T^b_\mu \\
&= \frac{1}{2^n} \partial_{\mu} \delta\Box^n T_{\nu}^{ a} - \frac{1}{2^n} \partial_{\nu} \delta \Box^n T_{\mu}^{ a} + \frac{1}{4^n} gf^{abc}  \Box^n T^b_\mu  \delta\Box^n T^c_\nu  -  \frac{1}{4^n} gf^{acb} \Box^n T^c_\nu \delta\Box^n T^b_\mu \\
&=\bigg( \frac{1}{2^n} \partial_{\mu} \delta\Box^n T_{\nu}^{ a} + \frac{1}{4^n} gf^{abc} \Box^n T^b_\mu  \delta \Box^n T^c_\nu  \bigg)- \bigg(\frac{1}{2^n} \partial_{\nu} \delta \Box^n T_{\mu}^{ a} + \frac{1}{4^n} gf^{abc} \Box^n T^b_\nu \delta\Box^n T^c_\mu  \bigg) \\
&:= \frac{1}{2^n}(D_{n\,\mu}\delta \Box^n T_{\nu}^a -D_{n\,\nu}\delta \Box^n T_{\mu}^a ) \,,
\end{aligned}
\end{equation}
where we define the covariant derivative of the rotor model as
\begin{equation} \label{eq:covariant}
D_{n\,\mu} =  \partial_{\mu} - \frac{i}{2^n} g\Box^n T_{\mu}^c t^{c}  \,,
\end{equation}
or in tensor form
\begin{equation}
D^{ab}_{n\,\mu} =  \partial_{\mu}\delta^{ab} - \frac{i}{2^n} g\Box^n T_{\mu}^c (t^{c})^{ab}  \,,
\end{equation}
and using the adjoint representation $(t^c)^{ab} = -if^{cab}$, then we have
\begin{equation} 
D^{ab}_{n\,\mu} =  \partial_{\mu}\delta^{ab} - \frac{1}{2^n} gf^{cab}\Box^n T_{\mu}^c  =  \partial_{\mu}\delta^{ab} - \frac{1}{2^n} gf^{abc}\Box^n T_{\mu}^c   \,.
\end{equation}
Therefore the covariant derivative in adjoint representation is 
\begin{equation} \label{eq:covariantderivative}
D^{ab}_{n\,\mu} = \partial_{\mu}\delta^{ab} + \frac{1}{2^n} gf^{acb}\Box^n T_{\mu}^c  \,.
\end{equation}
Therefore we have
\begin{equation}
\begin{aligned}
\frac{1}{2^n}D_{n\,\mu}^{ab}\delta \Box^n T_{\nu}^b = \frac{1}{2^n}D_{n\,\mu}\delta \Box^n T_{\nu}^a &=\frac{1}{2^n}\bigg( \partial_{\mu}\delta^{ab} + \frac{1}{2^n} gf^{acb}\Box^n T_{\mu}^c  \bigg)\delta \Box^n T_{\nu}^b \\
&=  \frac{1}{2^n} \partial_{\mu} \delta\Box^n T_{\nu}^{ a} + \frac{1}{4^n} gf^{abc}  \Box^n T^b_\mu \delta \Box^n T^c_\nu 
\end{aligned}
\end{equation}
as expected. Similarly,
\begin{equation}
\frac{1}{2^n}D_{n\,\nu}^{ab}\delta \Box^n T_{\mu}^b =  \frac{1}{2^n} \partial_{\nu} \delta\Box^n T_{\mu}^{ a} + \frac{1}{4^n} gf^{abc}  \Box^n T^b_\nu \delta \Box^n T^c_\mu 
\end{equation}
as expected. Now the variation of action becomes
\begin{equation}
\begin{aligned}
\delta S_{\mathrm{YM}}^{(n)} &= -\frac{1}{2} \int d^D x G_n^{\mu\nu a} \delta G_{n\,\mu\nu}^a \\
&= -\frac{1}{2}\int d^D x G_n^{\mu\nu a}(D_{n\,\mu}\delta \Box^n T_{\nu}^a -D_{n\,\nu}\delta \Box^n T_{\mu}^a   ) \\
&= \int d^D x G_n^{\mu\nu a} D_{n\,\mu} \delta\Box^n T_{\nu}^a \\
&=-\int d^D x (D_{n \,\mu} G_n^{\mu\nu a}) \delta \Box^n T_{\nu}^a \,,
 \end{aligned}
\end{equation}
where from the third line to the forth line we have performed integration by parts. The action is extremeized when
\begin{equation}
\frac{\delta S_{\mathrm{YM}}^{(n)}}{\delta \Box^n T_{\nu}^a} =0 \,.
\end{equation}
Thus we obtain the equation of motion for the generalized non-abelian gauge field theory as follow
\begin{equation}
D_{n \,\mu} G_n^{\mu\nu a} = 0 \,,
\end{equation}
or using equation (\ref{eq:covariantderivative}), acting on $G_{\mu\nu}^b$
\begin{equation} \label{eq:finalEOM}
 \partial_{\mu} G^{\mu\nu a}_n + \frac{1}{2^n} gf^{abc}\Box^n T_{\mu}^b G^{\mu\nu c}_n =0 \,.
\end{equation}
Explicitly expanding equation (\ref{eq:finalEOM}), the original equation of motion becomes (\ref{eq:expandEOM}) under the rotor mechanism, which is as follow,
\begin{equation}
\begin{aligned}
2\hat{R}^{\mu\nu}\Box^n T_\mu^a +gf^{abc} &\Big( \frac{1}{2^n}(2\Box^n T_{\mu}^a \partial^{\mu}\Box^n T^{\nu c} + (\partial_{\mu} \Box^n T^{\mu b}) \Box^n T^{\nu c} + \Box^n T_{\mu}^b \partial^\nu \Box^n T^{\mu k})\\
&\quad\quad\quad\quad \quad\quad \quad\quad \quad\quad \quad\quad    + \frac{1}{8^n}gf^{cde} \Box^n T_{\mu}^b \Box^n T^{\mu d}\Box^n T^{\nu e}\Big) =0\,.
\end{aligned}
\end{equation}
Under the Lorentz gauge condition, the equation of motion simplifies to
\begin{equation}
2\Box^{n+1} T^{\nu a} +gf^{abc} \Big( \frac{1}{2^n}(2\Box^n T_{\mu}^a \partial^{\mu}\Box^n T^{\nu c} + \Box^n T_{\mu}^b \partial^\nu \Box^n T^{\mu k}) + \frac{1}{8^n}gf^{cde} \Box^n T_{\mu}^b \Box^n T^{\mu d}\Box^n T^{\nu e}\Big) =0\,.
\end{equation}
The general covariant derivative of the rotor model in (\ref{eq:covariant}) in matrix form satisfies the Jacobi identity,
\begin{equation}
[D_{n\,\rho} , [D_{n\,\mu}, D_{n\,\nu}]] +[D_{n\,\mu} , [D_{n\,\nu}, D_{n\,\rho}]] + [D_{n\,\nu} , [D_{n\,\rho}, D_{n\,\mu}]] =0 \,.
\end{equation}
And since the commutator of covariant derivative is promoted to
\begin{equation}
[D_{n\,\mu}, D_{n\,\nu}] = -ig F_{n\,\mu\nu} \,,
\end{equation}
therefore we obtain the Bianchi identity of high-order covariant derivative as
\begin{equation}
D_{n\,\rho} F_{n\,\mu\nu}^a + D_{n\,\mu} F_{n\,\nu\rho}^a + D_{n\,\nu} F_{n\,\rho\mu}^a = 0 \,.
\end{equation}

Finally, we will compute the Noether's current and the associated Noether's charge of our high-order non-abelian gauge field theory under rotor model. First we know that by the transformation of field due to the rotor mechanism $T_{\mu}^a \rightarrow \Box^n T_{\mu}^a $, the Euler-Lagrangian equation becomes,
\begin{equation} \label{eq:EL}
\frac{\partial \mathcal{L}^{(n)}_{\mathrm{YM}}}{\partial \Box^n T_{\nu}^a} =   \partial_{\mu} \frac{\partial \mathcal{L}^{(n)}_{\mathrm{YM}}}{\partial ( \partial_{\mu} \Box^n T_{\nu}^a )} \,.
\end{equation}
Using the Euler-Lagrangian equation, it can be shown that by the same derivation as in \citep{BW}(equation 103), that the Noether current is identified by
\begin{equation}
 J^{\mu} =  \frac{\partial \mathcal{L}^{(n)}_{\mathrm{YM}}}{\partial ( \partial_{\mu} \Box^n T^a_{\nu} )} \delta \Box^n T^a_{\nu} \,.
\end{equation}
The Lagrangian can be broken down into three separate parts: the Maxwell part, the mixing part and the self-coupling part.
\begin{equation}
\mathcal{L}^{(n)}_{\mathrm{YM}} =\mathcal{L}^{(n)}_{\mathrm{Maxwell}} + \mathcal{L}^{(n)}_{\mathrm{mixing}} + \mathcal{L}^{(n)}_{\mathrm{self-coupling}}\,.
\end{equation}
The Maxwell part is in $2n$-order derivatives,
\begin{equation}
\mathcal{L}^{(n)}_{\mathrm{Maxwell}}=\frac{1}{4^n} \Box^n T^{\mu a} \hat{R}_{\mu\nu} \Box^n T^{\nu a} = -\frac{1}{4^{n+1}} \Box^n \tilde{G}_{\mu\nu}^a \Box^n \tilde{G}^{\mu\nu a}\,.
\end{equation} 
The mixing part is in $n$-order derivatives,
\begin{equation}
 \mathcal{L}^{(n)}_{\mathrm{mixing}} = -\frac{1}{2^{3n+1}}g f^{abc}(\partial^{\mu}\Box^n T^{\nu a} - \partial^{\nu} \Box^n T^{\mu a})\Box^n T_{\mu}^b \Box^n T_{\nu}^c \,.
\end{equation}
The self-coupling part is in 0-order derivatives,
\begin{equation}
\mathcal{L}^{(n)}_{\mathrm{self-coupling}}= -\frac{g^2}{2^{4n+2}} f^{abc}f^{ade} \Box^n T_{\mu}^b \Box^n T_{\nu}^c \Box^n T^{\mu d} \Box^n T^{\nu e} \,.
\end{equation}
Hence the Noether current is
\begin{equation}
J^\mu =\bigg(   \frac{\partial \mathcal{L}^{(n)}_{\mathrm{Maxwell}}}{\partial ( \partial_{\mu} \Box^n T^a_{\nu} )} +  \frac{\partial \mathcal{L}^{(n)}_{\mathrm{mixing}}}{\partial ( \partial_{\mu} \Box^n T^a_{\nu} )} +  \frac{\partial \mathcal{L}^{(n)}_{\mathrm{self-coupling}}}{\partial ( \partial_{\mu} \Box^n T^a_{\nu} )}   \bigg) \delta \Box^n T^a_{\nu}  \,.
\end{equation}
The first term is the standard Maxwell case, which is evaluated as
\begin{equation}
 \frac{\partial \mathcal{L}^{(n)}_{\mathrm{Maxwell}}}{\partial ( \partial_{\alpha} \Box^n T^k_{\beta} )} = -\frac{1}{4^n}\Box^n \tilde{G}^{\alpha\beta \,k} \,.
\end{equation}
The second term is evaluated to be
\begin{equation}
\begin{aligned}
 \frac{\partial \mathcal{L}^{(n)}_{\mathrm{mixing}}}{\partial ( \partial_{\alpha} \Box^n T^k_{\beta} )} &= \frac{\partial}{\partial ( \partial_{\alpha} \Box^n T^k_{\beta} )}  \frac{-1}{2^{3n+1}}g f^{abc}(\partial^{\mu}\Box^n T^{\nu a} - \partial^{\nu} \Box^n T^{\mu a})\Box^n T_{\mu}^b \Box^n T_{\nu}^c \\
 &=-\frac{1}{2^{3n+1}} gf^{abc}\bigg( \frac{\partial (\partial^{\mu}\Box^n T^{\nu a})}{\partial(\partial_{\alpha} \Box^n T_{\beta}^k )} - \frac{\partial (\partial^{\nu}\Box^n T^{\mu a})}{\partial(\partial_{\alpha} \Box^n T_{\beta}^k )} \bigg) \Box^n T_{\mu}^b \Box^n T_{\nu}^{c} + 0 \\
 &=-\frac{1}{2^{3n+1}}g f^{abc}(\delta^{\mu\alpha}\delta^{\nu\beta}\delta^a_k - \delta^{\nu\alpha}\delta^{\mu\beta}\delta^a_k  )  \Box^n T_{\mu}^b \Box^n T_{\nu}^{c} \\
 &= -\frac{1}{2^{3n+1}}g f^{kbc} ( \Box^n T^{\alpha b} \Box^n T^{\beta c} - \Box^n T^{\alpha c} \Box^n T^{\beta b}  ) \\
 &= -\frac{1}{2^{3n+1}}g f^{kbc} \Box^n T^{\alpha [b}\Box^n T^{\beta\,\,,c]} \,.
\end{aligned}
\end{equation}
The third term is zero as there are no derivatives of the rotor field,
\begin{equation}
\frac{\partial \mathcal{L}^{(n)}_{\mathrm{self-coupling}}}{\partial ( \partial_{\mu} \Box^n T^a_{\nu} )} =0 \,.
\end{equation}
Therefore, the Noether's current is
\begin{equation}
J^{\alpha} =\bigg(-\frac{1}{4^n}\Box^n \tilde{G}^{\alpha\beta \,k} -\frac{1}{2\cdot 8^n}g f^{kbc} ( \Box^n T^{\alpha b} \Box^n T^{\beta c} - \Box^n T^{\alpha c} \Box^n T^{\beta b}  )\bigg)\delta \Box^n T^k_{\beta} \,.
\end{equation}
The associated Noether's charge is given by
\begin{equation}
Q=\int d^{D-1}x j^0 = \int d^{D-1}x\bigg(-\frac{1}{4^n}\Box^n \tilde{G}^{0\beta \,k} -\frac{1}{2\cdot 8^n}g f^{kbc} ( \Box^n T^{0 b} \Box^n T^{\beta c} - \Box^n T^{0 c} \Box^n T^{\beta b}  )\bigg)\delta \Box^n T^k_{\beta} \,.
\end{equation}

\section{The issue of dynamic instability of generalized Yang-Mills theory under rotor mechanism  }
It is well known that quantum field theory with high-order derivative suffers from dynamic instability, in which the canonical Hamiltonian (energy) is unbounded below \cite{ho6e,ho6f,ho6g,ho6h}. The idea of dynamic instability aroused from high-order derivative systems was first established by Ostrogradsky, which is famously known as the Ostrogradsky theorem \cite{ho6d}. In this section, we will first give the Ostrogradsky construction of generalized Yang-Mills theory under rotor mechanism, then study the 00-component of the energy-momentum tensor which is regarded as the energy density of the system. Generally, if the 00-component of the energy-momentum tensor $T^{\mu}_{\,\,\,\nu}$ is greater or equal than 0, i.e. $T^0_{\,\,\,0} \geq 0$, the system is bounded and still considered as stable \cite{new1,new2,new3,new4}, even though the canonical energy is unbounded below. In other words, the energy is given by
\begin{equation}
E = \int d^{D-1} x \, T^0_{\,\,\,0} 
\end{equation} 
is greater or equal than zero, which is bounded.

Now we will first give an review of Ostrogradsky construction, then apply it to our case of generalized Yang-Mills theory. Consider a Lagrangian with higher-order time derivatives up to $n$, $L(x, \dot{x}, \ddot{x} ,\cdots, x^{(n)} )$, the Euler-Lagrange equation (equation of motion) reads \cite{first,ho6c,ho6e,new5},
\begin{equation}
\frac{\partial L}{\partial x} - \frac{d}{dt} \frac{\partial L}{\partial \dot{x}} + \frac{d^2}{dt^2} \frac{\partial L}{\partial \ddot{x}} + \cdots + (-1)^n \frac{d^n}{dt^n} \frac{\partial L}{\partial x^{(n)}} = \sum_{i=0}^n \bigg(-\frac{d}{dt} \bigg)^i \frac{\partial L}{\partial x^{(i)}} = 0 \,.
\end{equation}
The canonical variables are defined as follow \cite{first,ho6c,ho6e,new5},
\begin{equation}
X_i = x^{(i-1)}\,\, , \quad\quad P_i = \sum_{j=i}^n \bigg( -\frac{d}{dt}  \bigg)^{j-i} \frac{\partial L}{\partial x^{(j)}} \,.
\end{equation}
The canonical Hamiltonian (energy) is then given by \cite{first,ho6c,ho6e,new5},
\begin{equation} \label{eq:OH}
H= \sum_{i=1}^m P_i x^{(i)} -L = P_1 X_2 + P_2 X_3 + \cdots + P_{n-1} X_n + P_n \dot{X}_n - L(X_1 , X_2 , \cdots , X_n ,  \dot{X}_n) \,.
\end{equation}
The Ostrogradsky Hamiltonian in equation (\ref{eq:OH}), which is known as the canonical energy, is linear in canonical momentum variable $P_i$, implying the energy system can be lowered without any bound. Hence, the inclusion of higher order derivatives make the system unstable. 

The Ostrogradsky construction for our generalized Yang-Mills theory under rotor mechanism is as follow. Consider the Lagrangian density as a functional of all orders of rotored fields $\mathcal{L}(A_{\nu} , \Box A_{\nu},  \Box^2 A_{\nu} ,\cdots , \, \Box^n A_\nu  )$, for example,
\begin{equation}
\mathcal{L}(A^a_{\nu} , \Box A^a_{\nu},  \Box^2 A^a_{\nu} , \cdots , \Box^n A_\nu^a  ) = -\frac{1}{4} \sum_{k=1}^n a_k G_{k\,\mu\nu}^a G_{k}^{\mu\nu a} \,,
\end{equation}
where $a_k$ is some coefficient. The Euler-Lagrangian equation is
\begin{equation}
\frac{\partial \mathcal{L}}{\partial A_{\nu}^a} + \Box \frac{\partial \mathcal{L}}{\partial \Box A_{\nu}^a} + \Box^2 \frac{\partial\mathcal{L}}{\partial \Box^2 A_{\nu}^a} + \cdots + \Box^n \frac{\partial\mathcal{L}}{\partial \Box^n A_\nu^a} = \sum_{k=0}^n \Box^k \bigg(\frac{\partial\mathcal{L}}{\partial \Box^k A_\nu} \bigg)= 0 \,.
\end{equation}
Note that there are no negative terms because the $\Box^k$ operator is in even order of derivative. The canonical variables are,
\begin{equation}
X_{i\nu} = \Box^{i-1} A_{\nu} \,\, , \quad\quad P_i^\nu = \sum_{j=i}^n \Box^{j-i} \frac{\partial \mathcal{L}}{\partial \Box^j A_\nu} \,.
\end{equation} 
The Ostrogradsky Hamiltonian, or the canonical energy is 
\begin{equation}
\mathcal{H} =\sum_{i=1}^n P_i^\nu \Box^i A_{\nu} - \mathcal{L} =\sum_{i=1}^n \sum_{j=i}^n \bigg(\Box^{j-i} \frac{\partial \mathcal{L}}{\partial \Box^j A_\nu}  \bigg)\Box^i A_{\nu} - \mathcal{L}  \,.
\end{equation}
Since the Hamiltonian is linear in $P_i^\nu$, it is unbounded below. Hence, the generalized Yang-Mills theory under rotor model is unstable in terms of the canonical energy.  

Next, we would evaluate the conserved energy-momentum tensor (also known as the stress-energy tensor). Consider the generic Yang-Mills action under rotor mechanism in curved spacetime,
\begin{equation} \label{eq:YMgeneric}
S^{(n)}_{\mathrm{YM}} = -\frac{1}{4}\int d^D x \sqrt{-g} G_{n\,\mu\nu}^a G_{n}^{\mu\nu\,a} \,,
\end{equation} 
where $g = \det g_{\mu\nu}$ is the determinant of the metric. The energy momentum tensor is given by
\begin{equation}
T^{\mu\nu}_{n} = -\frac{2}{\sqrt{-g}} \frac{\delta S^{(n)}_{\mathrm{YM}}}{\delta g_{\mu\nu}} \,.
\end{equation}
This gives the energy-momentum tensor of generalized Yang-Mills theory under the rotor model as
\begin{equation}
T^{\mu\nu}_n = G^{\mu\lambda\,a}_n G^{\nu\,a}_{n \lambda} -\frac{1}{4} g^{\mu\nu}G_{n}^{\alpha\beta\,a}G_{n\,\alpha\beta}^a \,.
\end{equation}
Alternatively, this can be obtained by using the modified Noether's theorem in reference \cite{new6}. Here as we are interested in the case of flat spacetime, we take $g_{\mu\nu} = \eta_{\mu\nu}$, therefore we have
\begin{equation}
T^{\mu\nu}_n = G^{\mu\lambda\,a}_n G^{\nu\,a}_{n\lambda} -\frac{1}{4} \eta^{\mu\nu}G_{n}^{\alpha\beta\,a}G_{n\,\alpha\beta}^a \,.
\end{equation}
In-terms of rank(1,1) tensor, we have
\begin{equation} \label{eq:stress1}
T^{\mu}_{n\,\,\,\nu} = G^{\mu\lambda\,a}_n G^{\,a}_{n\nu\lambda} -\frac{1}{4} \delta^{\mu}_{\,\,\,\nu}G_{n}^{\alpha\beta\,a}G_{n\,\alpha\beta}^a \,.
\end{equation}
The total action is given by the sum of the action in (\ref{eq:YMgeneric}) for all orders, i.e.
\begin{equation}
S_{\mathrm{YM}\,tot} = \sum_{k=0}^{n} a_k S^{(k)}_{\mathrm{YM}} = -\frac{1}{4}\int d^D x \sum_{k=0}^n \,  a_k G_{k\,\mu\nu}^a G_{k}^{\mu\nu\,a} \,.
\end{equation}
The total energy-momentum tensor is
\begin{equation}
(T^{\mu}_{\,\,\,\nu})_{tot} = \sum_{k=0}^n \beta_k T^{\mu}_{k\,\,\nu} \,,
\end{equation}
where $ \beta_k$ is the linear coefficient of the corresponding energy-momentum tensor. As we are interested in the 00-component, which is the energy density, therefore the total energy density of the system is
\begin{equation}
\mathcal{E}_{tot} =  (T^{0}_{\,\,\,0})_{tot} = \sum_{k=0}^n \beta_k T^{0}_{k\,\,0} \,.
\end{equation}
The explicit evaluation of $T^{0}_{n\,\,0}$ for the non-abelian case is rather complicated. It is more convenient to study its abelian counterpart first such that we get a brief idea, and return to the non-abelian case afterwards. 

For the abelian case, the gauge symmetry is U(1) and there is only one generator. For further simplicity we work for the $n=0$ case first, which is the original Maxwell theory. The Maxwell stress-energy tensor in (\ref{eq:stress1}) reduces to
\begin{equation}
T^{\mu}_{0\,\,\,\nu} = G^{\mu\lambda} G_{\nu\lambda} -\frac{1}{4} \delta^{\mu}_{\,\,\,\nu}G^{\alpha\beta}G_{\alpha\beta} \,.
\end{equation} 
By explicitly expanding the gauge field strength tensor of the first term, and carry out integration by parts of the last term \footnote{When we calculate the actual energy-momentum tensor (not density), we have $\tau^{\mu}_{\,\,\,\nu} = \int d^{D-1}x T^{\mu}_{\,\,\,\nu}$, so the presence of the integral allows us to do integration by parts.} , we have
\begin{equation}
\begin{aligned}
T^{\mu}_{0\,\,\,\nu} &= (\partial^\mu T^\lambda - \partial^\lambda T^\mu)( \partial_\nu T_\lambda  - \partial_\lambda T_\nu  ) + \delta^{\mu}_{\,\,\,\nu} T^{\alpha}\hat{R}_{\alpha\beta} T^\beta \\
&= \partial^\mu T^\lambda \partial_{\nu}T_\lambda -\partial^\mu T^\lambda \partial_\lambda T_\nu - \partial^\lambda T^\mu \partial_\nu T_\lambda + \partial^\lambda T^\mu \partial_\lambda T_\nu  + \delta^{\mu}_{\,\,\,\nu} T^{\alpha}\hat{R}_{\alpha\beta} T^\beta \,.
\end{aligned}
\end{equation}
The 00-component is
\begin{equation}
\begin{aligned}
T^{0}_{0\,\,\,0} &= \partial^0 T^\lambda \partial_{0}T_\lambda -\partial^0 T^\lambda \partial_\lambda T_0 - \partial^\lambda T^0 \partial_0 T_\lambda + \partial^\lambda T^0 \partial_\lambda T_0  +  T^{\alpha}\hat{R}_{\alpha\beta} T^\beta \\
&=\partial^0 T^\lambda \partial_{0}T_\lambda -\partial^0 T^\lambda \partial_\lambda T_0 - \partial_\lambda T_0 \partial^0 T^\lambda + \partial^\lambda T^0 \partial_\lambda T_0  +  T^{\alpha}\hat{R}_{\alpha\beta} T^\beta \\
&= \partial^0 T^\lambda \partial_{0}T_\lambda - 2\partial^0 T^\lambda \partial_\lambda T_0  + \partial^\lambda T^0 \partial_\lambda T_0  +  T^{\alpha}\hat{R}_{\alpha\beta} T^\beta \,.
\end{aligned}
\end{equation}
As we use the $+---$ convention of the metric, we have $\partial_0 = \partial^0 = \frac{\partial}{\partial t}$, and $A_0 = A^0 = V$ is the potential, we have
\begin{equation}
\begin{aligned}
T^{0}_{0\,\,\,0} &= \frac{\partial T^\lambda}{\partial t} \frac{\partial T_\lambda}{\partial t} -2 \frac{\partial T^\lambda}{\partial t} \partial_\lambda V + \partial^\lambda V \partial_\lambda V +  T^{\alpha}\hat{R}_{\alpha\beta} T^\beta \\
&\equiv \dot{T}^\lambda \dot{T}_\lambda -2\dot{T}^\lambda \partial_\lambda V +\partial^\lambda V \partial_\lambda V +  T^{\alpha}\hat{R}_{\alpha\beta} T^\beta \\
&= ( \dot{T}^\lambda -\partial^\lambda V       )^2 + T^{\alpha}\hat{R}_{\alpha\beta} T^\beta \,.
\end{aligned}
\end{equation}
Since the first term is a quadratic term, it is always greater or equal than zero, and the second term is quadratic in nature, we have $T^{0}_{0\,\,\,0} \geq 0$. The energy density for the Maxwell theory is bounded. 

Now we consider the energy momentum tensor of the generalized abelian gauge field theory under rotor model. Under the rotor mechanism, the gauge field transforms as $G_{n \mu\nu} = \frac{1}{2^n} \Box^n G_{\mu\nu}$ \cite{BW}, therefore we have
\begin{equation}
T^{\mu}_{n\,\,\,\nu} = G^{\mu\lambda}_n G_{n\nu\lambda} -\frac{1}{4} \delta^{\mu}_{\,\,\,\nu}G_{n}^{\alpha\beta}G_{n\,\alpha\beta} = \frac{1}{4^n} \bigg( \Box^n G^{\mu\lambda} \Box^n G_{\nu\lambda} -\frac{1}{4} \delta^{\mu}_{\,\,\,\nu} \Box^n G^{\alpha\beta}  \Box^n G_{\alpha\beta}  \bigg) \,.
\end{equation} 
The 00-component is
\begin{equation}
\begin{aligned}
T^{0}_{n\,\,\,0} &= \frac{1}{4^n}\bigg(\frac{\partial \Box^n T^\lambda}{\partial t} \frac{\partial \Box^n T_\lambda}{\partial t} -2\frac{\partial \Box^n T^\lambda}{\partial t} \partial_{\lambda} \Box^n V + \partial^\lambda \Box^n V \partial_{\lambda}\Box^n V + \Box^n T^\alpha \hat{R}_{\alpha\beta}\Box^n T^\beta   \bigg) \\
&=  \frac{1}{4^n} \bigg[ \bigg( \frac{\partial \Box^n T^\lambda}{\partial t} - \partial^\lambda \Box^n V   \bigg)^2 + \Box^n T^{\alpha} \hat{R}_{\alpha\beta} \Box^n T^{\beta} \bigg] \,.
\end{aligned}
\end{equation}
Since both terms are quadratic in nature, therefore we must have $T^{0}_{n\,\,\,0} \geq 0 $ for all $n$. Hence, the energy density of the rotor model is bounded. When $n \rightarrow \infty$, we have $T^{0}_{\infty\,\,\,0} =0$, which is the lower limit. Now we have the total energy density as
\begin{equation}
\mathcal{E}_{tot}  =  (T^{0}_{\,\,\,0})_{tot}= \sum_{k=0}^n \beta_k \frac{1}{4^k} \bigg[ \bigg( \frac{\partial \Box^k T^\lambda}{\partial t} - \partial^\lambda \Box^k V   \bigg)^2 + \Box^k T^{\alpha} \hat{R}_{\alpha\beta} \Box^k T^{\beta} \bigg]
\end{equation}
If $\beta_k \geq 0$ for all $k$, it is guaranteed that $\mathcal{E}_{tot} \geq 0 $. 

Finally, we proceed to work for the general non-abelian case. Starting from equation (\ref{eq:stress1}), and using the result in (\ref{eq:transform}),  we expand explicitly to give
\begin{equation}
\begin{aligned}
T^{\mu}_{n \,\, \nu} &= (\tilde{G}^{\mu\lambda\,a}_n + \frac{1}{4^n} g f^{abc} \Box^n T^{\mu b } \Box^n T^{\lambda c}   )( \tilde{G}^{\,a}_{n\nu\lambda} + \frac{1}{4^n} f^{ade}\Box^n T^d_\nu \Box^n T^e_\lambda ) \\
& \,\,\,\,\,\,\,\,+\delta^{\mu}_{\nu} \bigg( \frac{1}{4^n} \Box^n T^{\alpha a} \hat{R}_{\alpha\beta} \Box^n T^{\beta a} - \frac{1}{2\cdot 4^n }g f^{abc} \tilde{G}^{\alpha\beta\,a}_n  \Box^n T_{\alpha}^b \Box^n T_{\beta}^c \\
&\quad\quad \quad\quad\quad - \frac{g^2}{4\cdot 16^n} f^{abc}f^{ade} \Box^n T_{\alpha}^b \Box^n T_{\beta}^c \Box^n T^{\alpha d} \Box^n T^{\beta e} \bigg) \\
&= \tilde{G}^{\mu\lambda\,a}_n \tilde{G}^{\,a}_{n\nu\lambda} + \frac{1}{4^n}gf^{ade} \tilde{G}^{\mu\lambda\,a}_n \Box^n T^d_\nu \Box^n T^e_\lambda + \frac{1}{4^n} gf^{abc} \tilde{G}^{\,a}_{n\nu\lambda} \Box^n T^{\mu b } \Box^n T^{\lambda c} \\
& \quad\quad + \frac{1}{16^n} g^2 f^{abc}f^{ade}\Box^n T^{\mu b } \Box^n T^{\lambda c} \Box^n T^d_\nu \Box^n T^e_\lambda \\
& \quad\quad+\delta^{\mu}_{\nu} \bigg( \frac{1}{4^n} \Box^n T^{\mu a} \hat{R}_{\mu\nu} \Box^n T^{\nu a} - \frac{1}{2\cdot 4^n }g f^{abc} \tilde{G}^{\mu\nu\,a}_n  \Box^n T_{\mu}^b \Box^n T_{\nu}^c \\
&\quad\quad \quad\quad\quad - \frac{g^2}{4\cdot 16^n} f^{abc}f^{ade} \Box^n T_{\mu}^b \Box^n T_{\nu}^c \Box^n T^{\mu d} \Box^n T^{\nu e} \bigg) \\
\end{aligned}
\end{equation}
Next we evaluate the $00$-component, 
\begin{equation}
\begin{aligned}
T^{0}_{n \,\, 0} & = \frac{1}{4^n} \bigg[ \bigg( \frac{\partial \Box^n T^{\lambda a}}{\partial t} - \partial^{\lambda a} \Box^n V^a   \bigg)^2 + \Box^n T^{\alpha a} \hat{R}_{\alpha\beta} \Box^n T^{\beta a} \bigg] \\
&\quad\quad +\frac{2}{4^n} g f^{abc} \tilde{G}^a_{n0\lambda}\Box^n V^b \Box^n T^{\lambda c} + \frac{1}{16^n} g^2 f^{abc}f^{ade}\Box^n V^b \Box^n V^d \Box^n T^{\lambda c} \Box^n T^e_\lambda \\
& \quad\quad - \frac{1}{2\cdot 4^n }g f^{abc} \tilde{G}^{\mu\nu\,a}_n  \Box^n T_{\mu}^b \Box^n T_{\nu}^c  - \frac{g^2}{4\cdot 16^n} f^{abc}f^{ade} \Box^n T_{\mu}^b \Box^n T_{\nu}^c \Box^n T^{\mu d} \Box^n T^{\nu e} 
\end{aligned}
\end{equation}
Then finally we obtain,
\begin{equation} \label{eq:finalresult}
\begin{aligned}
T^{0}_{n \,\, 0} & = \frac{1}{4^n} \bigg[ \bigg( \frac{\partial \Box^n T^{\lambda a}}{\partial t} - \partial^{\lambda} \Box^n V^a   \bigg)^2 + \Box^n T^{\alpha a} \hat{R}_{\alpha\beta} \Box^n T^{\beta a} \bigg] \\
&\quad\quad+\frac{2}{4^n}gf^{abc}\bigg(  \tilde{G}^a_{n0\lambda}\Box^n V^b \Box^n T^{\lambda c} -\frac{1}{4} \tilde{G}^{\mu\nu\,a}_n \Box^n T^b_\mu \Box^n T^c_\nu    \bigg) \\
&\quad\quad + \frac{g^2}{16^n} f^{abc}f^{ade}\bigg(\Box^n V^b \Box^n V^d \Box^n T^{\lambda c} \Box^n T^e_\lambda -\frac{1}{4} \Box^n T_{\mu}^b \Box^n T_{\nu}^c \Box^n T^{\mu d} \Box^n T^{\nu e}    \bigg)
\end{aligned}
\end{equation}
The first line of the final equation (\ref{eq:finalresult}) is Maxwellian-like, which is greater or equal than zero. However, the non-linear terms provided by the structural constant are not necessarily positive, unless the following constraint is satisfied,
\begin{equation} \label{eq:one}
\tilde{G}^a_{n0\lambda}\Box^n V^b \Box^n T^{\lambda c} -\frac{1}{4} \tilde{G}^{\mu\nu\,a}_n \Box^n T^b_\mu \Box^n T^c_\nu \geq 0 
\end{equation} 
and
\begin{equation} \label{eq:two}
\Box^n V^b \Box^n V^d \Box^n T^{\lambda c} \Box^n T^e_\lambda -\frac{1}{4} \Box^n T_{\mu}^b \Box^n T_{\nu}^c \Box^n T^{\mu d} \Box^n T^{\nu e} \geq 0 \,.
\end{equation}
The only certain thing is when $n\rightarrow \infty$, $T^0_{\infty\,\,0} =0$ is bounded. And for the total energy density, we demand all $\beta_k \geq 0$. Therefore, while the abelian case is bounded, there is no garantee for the non-abelian case is also bounded, unless the constraints by (\ref{eq:one}) and (\ref{eq:two}) are satisfied. This marks the difference between the abelian case and the non-abelian case.

\section{Conclusion}
In this paper, we have established the generalized non-abelian gauge field theorem under the rotor mechanism. Under the Lorentz gauge condition, the rotor transformation of gauge field for the general non-abelian case is same as the abelian case. The gauge field transforms as $T_{\mu}^a \rightarrow \Box^n T_{\mu}^a $ under the rotor mechanism. When $n=0$. this restores back to the original Yang-Mills theory. We also compute the equation of motion and Noether's current for our theory. Finally, we study the dynamic stability issue of both the abelian case and the non-abelian (Yang-Mills) case. Although the canonical energy is unbounded below, the 00-component of the energy-momentum tensor is still positive for the abelian case and thus can still considered as stable. However, the non-abelian case is much more complicated and the system is considered stable only if a certain criteria in the non-linear terms is satisfied. In both case when the rotor order $n$ is large enough and tends to infinity, the 00-component of the energy-momentum tensor is bounded to zero. In the future, this theory can help to develop the generalized field theory of higher order derivatives for the standard model of particles.

\subsubsection*{Declaration}
I declare that there are no known competing financial interests or personal relationships that could have appeared to influence the work reported in this paper.


\begin{thebibliography}{}
%
%
\bibitem{first}
A. Pais and G.E. Uhlenbeck. On field theories with non-localized action, \textit{Phys. Rev.} \textbf{ 79}. 145-165. 1980.

\bibitem{ho1}
B. Podolsky. A Generalized Electrodynamics Part I—Non-Quantum. \textit{Phys. Rev.} 62, \textbf{68}. 1942.

\bibitem{ho2}
B. Podolsky and C. Kikuchi.  A Generalized Electrodynamics Part II-Quantum. \textit{Phys. Rev.} \textbf{65}. 228. 1944.

\bibitem{ho3}
B. Podolsky and C. Kikuchi. Auxiliary Conditions and Electrostatic Interaction in Generalized Quantum Electrodynamics. \textit{Phys. Rev}. \textbf{67}, 184. 1945.

\bibitem{ho4}
B. Podolsky and P. Schwed. Review of a Generalized Electrodynamics. \textit{Rev. Mod. Phys}. 20, \textbf{40}. 1948.

\bibitem{ho5}
D. J. Montgomery. Relativistic Interaction of Electrons on Podolsky's Generalized Electrodynamics. \textit{Phys. Rev}.  \textbf{69}. 117. 1946.

\bibitem{LeeWick1}
T. D. Lee and G. C. Wick \textit{Nucl.Phys.B} \textbf{ 9}, 209. 1969.

\bibitem{LeeWick2}
T. D. Lee and G. C. Wick. \textit{Phys.Rev.D} \textbf{2}, 1033. 1970.

\bibitem{LeeWickSM}
B. Grinstein, D. O'Connell, M. B. Wise. The Lee-Wick Standard Model. \textit{Phys.Rev.D} \textbf{77}:025012. 2008.


\bibitem{ho6a}
G.W. Gibbons, C.N. Pope and Sergey Solodukhin. Higher Derivative Scalar Quantum Field Theory in Curved Spacetime. \textit{Phys.Rev.D} \textbf{100}. 2019. 


\bibitem{ho6b}
D.S. Kaparulin, S.L. Lyakhovich, O.D. Nosyrev, Extended Chern-Simons model for a vector multiplet, Symmetry 2021, 13(6) 1004.


\bibitem{ho6c}
D.S. Kaparulin. A stable higher-derivative theory with the Yang-Mills gauge symmetry.  	arXiv:2011.12928 [hep-th]



\bibitem{ho6d}
M. Ostrogradsky. Mem. Ac. St. Petersbourg VI 4 (1850) 385.

\bibitem{ho6e}
R. P. Woodard. The Theorem of Ostrogradsky. arXiv:1506.02210 [hep-th]. 

\bibitem{ho6f}
V.V. Nesterenko. On the instability of classical dynamics in theories with higher derivatives, \textit{Phys.Rev.D} \textbf{75}. 2007.

\bibitem{ho6g}
N.G. Stephen. On the Ostrogradski instability for higher-order derivative theories and apseudo-mechanical energy. \textit{J. Sound. Vib} \textbf{310}(3): 729-739ïijŇ. 2008.

\bibitem{ho6h}
H. Motohashi and T. Suyama. Third order equations of motion and the Ostrogradsky instability, \textit{Phys.Rev.D} \textbf{91}. 2015.



\bibitem{h1}
K. S. Stelle. Renormalization of higher-derivative quantum gravity. \textit{Phys.Rev.D} \textbf{16}. 953. 1977.

\bibitem{h2}
E. S. Fradkin and A.A. Tseytlin. Renormalizable asymptotically free quantum theory of gravity. \textit{Nucl. Phys.B} 201. 1982. 469.


\bibitem{h3}
S.Nojiri and S.D.Odintso. Introduction to Modified Gravity and Gravitational Alternative for Dark Energy. \textit{Int.J.Geom.Meth.Mod.Phys.}\textbf{4}. 2007. 

\bibitem{h4}
T.P. Sotiriou. $f(R)$ Theories of Gravity. \textit{Rev. Mod. Phys.}\textbf{ 82}, 451-497. 2010.

\bibitem{h5}
S. Nojiri and S. D. Odintsov.  \textit{Phys.Rept}.505. 2011. 59.

\bibitem{BW}
B.T.T.Wong. Generalized abelian gauge field theory under rotor model. \textit{Mod. Phys. Lett. A.} Vol. 36, No. 27, 2150194. 2021. 

\bibitem{YangMills}
Yang, C. N.; Mills, R. Conservation of Isotopic Spin and Isotopic Gauge Invariance. \textit{Physical Review}. 96 (1): 191–195. 1954

\bibitem{Peskin}
M. E. Peskin and  D.V. Schroeder. An introduction to quantum field theory. \textit{ABP}. 1995. 

\bibitem{new1}
J. Dai. Stability in the higher derivative Abelian gauge field theory. \textit{Nuclear Physics B}. Vol 961. 2020.


\bibitem{new2}
D. S. Kaparulin, S. L. Lyakhovich, A. A. Sharapov. Classical and quantum stability of higher-derivative dynamics. \textit{Eur. Phys. J. C}  \textbf{74}. 3072. 2014.


\bibitem{new3}
D. S. Kaparulin, I. Yu. Karataeva, S. L. Lyakhovich. Higher derivative extensions of 3d Chern-Simons models: conservation laws and stability. \textit{ Eur. Phys. J. C} \textbf{75}. 552. 2015. 


\bibitem{new4}
V.A. Abakumova, D.S. Kaparulin, S.L. Lyakhovich. Multi-Hamiltonian formulations and stability of higher-derivative extensions of 3d Chern-Simons. \textit{ The EPJ C} \textbf{ 78}. 115. 2018.


\bibitem{new5}
F.J. de Urries and J.Julve. Ostrogradski Formalism for Higher-Derivative Scalar Field Theories. \textit{J.Phys.A} \textbf{31}. 6949-6964. 1998.

\bibitem{new6}
M. Montesinos and E. Flores. Symmetric energy-momentum tensor in Maxwell, Yang-Mills, and Proca theories obtained using only Noether's theorem. \textit{Rev.Mex.Fis.} \textbf{52}. 29-36. 2006.  


\end{thebibliography}
\end{document}